\documentclass[aps,prl,twocolumn,amsmath,showpacs]{revtex4-1}
\usepackage{graphicx,dcolumn,bm,amssymb,amsmath,ulem,indentfirst,amsthm}
\usepackage[toc,page]{appendix}
\usepackage{color}

\theoremstyle{plain}
\def\be{\begin{equation}}
\def\ee{\end{equation}}

\newtheorem*{theorem*}{Theorem}

\begin{document}

\author{Bingyu Cui$^{1}$, Alessio Zaccone$^{1,2}$}
\affiliation{${}^1$Statistical Physics Group, Department of Chemical
Engineering and Biotechnology, University of Cambridge, Philippa Fawcett Drive, CB3
0AS Cambridge, U.K.}
\affiliation{${}^2$Cavendish Laboratory,University of Cambridge, JJ Thomson
Avenue, CB30HE Cambridge,
U.K.}
\begin{abstract}
The Generalized Langevin Equation (GLE) can be derived from a particle-bath
Hamiltonian, in both classical and quantum dynamics, and provides a route to
the (both Markovian and non-Markovian) fluctuation-dissipation theorem (FDT).
All previous studies have focused either on particle-bath systems with
time-independent external forces only, or on the simplified case where only the
tagged particle is subject to the external time-dependent oscillatory field.
Here we extend the GLE and the corresponding FDT for the more general case
where both the tagged particle and the bath oscillators respond to an external
oscillatory field. This is the example of a charged or polarisable particle
immersed in a bath of other particles that are also charged or polarizable,
under an external AC electric field.
For this Hamiltonian, we find that the ensemble average of the stochastic force
is not zero, but proportional to the AC field. The associated FDT reads as
$\langle F_P(t)F_P(t')\rangle=mk_BT\nu(t-t')+(\gamma e)^2E(t)E(t')$, where
$F_{p}$ is the random force, $\nu(t-t')$ is the friction memory function, and
$\gamma$ is a numerical prefactor.
\end{abstract}

\title{Generalized Langevin Equation and fluctuation-dissipation
theorem for particle-bath systems in external oscillating fields}
\maketitle

The theory of Brownian motion is so far the simplest approximation to the
dynamics of nonequilibrium systems. The Langevin equation, and its equivalent
counterpart, the Smoluchowski diffusion equation, describe the motion of a
Brownian particle in an external force field and under the action of thermal
agitation from the bath of solvent molecules, in which the Brownian particle is
immersed. At steady-state, the thermal fluctuations that cause rapid changes in
the particle velocity are dissipated by viscous drag. \\

This is a manifestation of the fluctuation-dissipation theorem (FDT), as was
originally formulated in different contexts by Einstein and by Nyquist, and
generalized by Onsager~\cite{Onsager} and by Callen and Welton \cite{Callen}.
Later, the theorem has been further elaborated in many different
contexts~\cite{Leontovich,Rytov,Kubo,Bernard,Akhiezer,Landau_el,Landau,Seifert1,Seifert2,Rubi}.
FDT stipulates that the response of a system in thermodynamic equilibrium to a
small applied force is the same as its response to a spontaneous fluctuation,
thus connecting the linear response relaxation of a system to equilibrium, from
a prepared nonequilibrium state, with its statistical fluctuation properties in
equilibrium. FDT applies to both classical and quantum mechanical
systems~\cite{Ford,Haenggi} and has been generalized to non-Markovian processes
for classical systems by Zwanzig~\cite{Zwanzig}. In the latter case, the noise
is no longer uncorrelated in time, and the time correlation of the stochastic
force is dictated by the time correlation of the friction which plays the role
of the memory function in the Generalized Langevin Equation (GLE). The
non-Markovianity arises from the dynamical coupling of the tagged Brownian
particle with many particles (harmonic oscillators) forming the heat bath. This
coupling, which in physical systems may be provided by long-range molecular
interactions, is thus responsible for both the thermal agitation and the
damping experienced by the tagged particle.
\\

All versions of the GLE, and of the associated FDT which can be derived from
it, that have been considered in the past are limited to either systems in the
absence of external time-dependent forces or, if an external time-dependent
force is present, its action is restricted to the tagged Brownian particle,
leaving the bath oscillators unaffected by the external field~\cite{Kubo,Zwerger}.
Harada and Sasa considered 
time-dependent driving forces being exerted on the tagged particle and they found a
violation of the fluctuation-response relation in driven nonequilibrium 
systems~\cite{Harada,Sasa}.
More recently, Maes and co-workers derived FDT for nonequilibrium systems by implementing 
mutual interactions between bath particles and also the effect of stochastic white-noise force on the bath particles dynamics ~\cite{Maes, Steffenoni}.  

This limitation is obviously artificial and not realistic, because in many
systems there is no reason to justify why the tagged particle is subjected to
the external AC field whereas the bath oscillators remain unaffected by the
field.
This limitation clearly leaves out a number of important physical problems,
where not only the tagged particle is subjected to the AC field, but also the
particles that constitute the heat bath are subjected to it.
This situation is clearly encountered in dielectric matter under a uniform AC
electric field $E(t)$. In this case, every building block (atom, molecule, ion)
is polarisable or charged such that it is subjected to a force from the
electric field. Hence, if the bath is constituted by polarisable or charged
particles, these will also respond to the AC field and it is unphysical to
neglect the effect of the AC field on the dynamics of the bath oscillators.
This situation is schematically depicted in Fig. 1. 

A similar situation arises in the field of
microrheology~\cite{Cicuta} where not only the probe Brownian particle, but also the building-blocks of the medium
do respond to the oscillatory mechanical field. 

In this contribution, we provide a solution to this problem by formulating a
Caldeira-Leggett particle-bath Hamiltonian where both the particle and the bath
oscillators are subjected to the external AC electric field, which is
explicitly accounted for in both the Hamiltonian of the particle and the
Hamiltonian of the bath. The two Hamiltonians are then connected via a
bi-linear coupling as is standard in this type of models.
We analytically solve the coupled Hamiltonian for the tagged particle to find a
new Generalized Langevin Equation, which, for the first time, accounts for the
effect of the polarization of the bath under the AC field on the dynamics of
the tagged particle. 

We also derive the associated FDT and find a surprising result: the
time-correlation of the stochastic force is not just equal to the memory
function for the friction, but there is an additional term which is
proportional to the time-correlation of the AC field, and hence to the
amplitude of the AC field squared. This term is absent in all previously
studied versions of the FDT. This result shows that the strength of the
fluctuating force can be controlled by the external force field in the limit of
sufficiently strong external driving.
This framework opens new avenues to understanding the macroscopic response of
complex liquids, plasmas, colloids and other condensed matter under oscillatory
fields.

\begin{figure}
\centering
\includegraphics[height=5cm ,width=8cm]{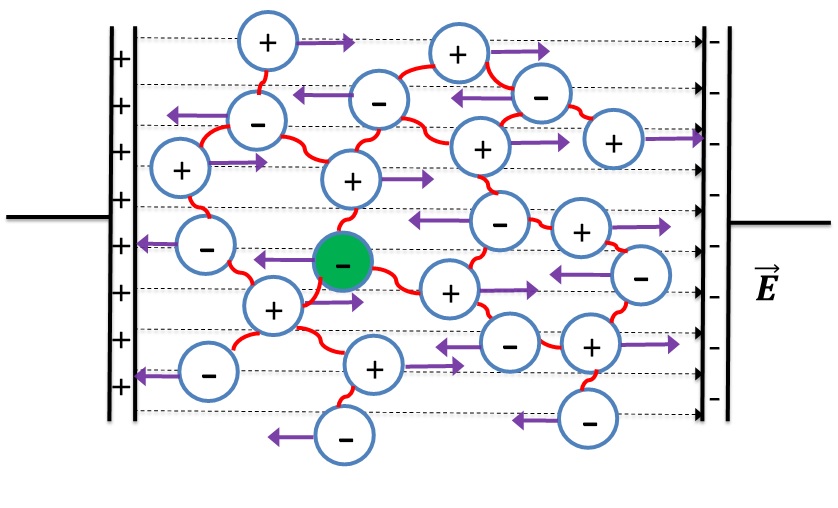}
\caption{Schematic example of system of charged (solid circles) or polarisable
molecules.
In the former case the particles could be ions in a plasma or ions and
electrons in a liquid metal. In the latter case, pairs of negatively and positively
charged particles represent the electron cloud and the molecular ion of a
polarised neutral molecule as in e.g. dielectric relaxation of molecular
liquids.
A particle-bath Hamiltonian like the one of Eqs.(1-2) can be applied to these
systems where a tagged particle (coloured, green) interacts with the local
environment via an interaction potential $V(Q)$, which may represent the
interaction potential with neighbours. The tagged particle interacts also with all other degrees of freedom in the
system which can be effectively represented as a bath of harmonic oscillators
to which the tagged particle is coupled via a set of coupling constants
$c_{\alpha}$, where $\alpha$ runs over all other bath oscillators in the
system. In traditional models of bath-oscillator dynamics, only the tagged
particle is subjected to the external AC field, see e.g. \cite{Kubo}, whereas
the other particles are not. In the proposed model, both the tagged particle
and also all the other particles (forming the bath) are responding to the AC
electric field, which leads to new physics.
}
\end{figure}

We study the classical version of the Caldeira-Leggett coupling between the
tagged particle and a bath of harmonic oscillators, which was actually proposed
already in 1973 by Zwanzig~\cite{Zwanzig}, and add a new term, $H_{E}$, which
contains the force due to the applied AC electric field acting on both the
tagged particle and the harmonic oscillators:

\begin{equation}
H=H_S+H_B+H_E
\end{equation}

where $H_S=P^2/2m+V(Q)$ is the Hamiltonian of the tagged particle without
external field. The second term is the standard Hamiltonian of the bath of
harmonic oscillators that are coupled to the tagged particle,
$H_B=\frac{1}{2}\sum_{\alpha=1}^N\left[\frac{P_{\alpha}^2}{m_{\alpha}}+m_{\alpha}\omega_{\alpha}^2\left(X_{\alpha}
-\frac{F_{\alpha}(Q)}{\omega_{\alpha}^2}\right)^{2}\right]$, consisting of the
standard harmonic oscillator expression for each bath oscillator $\alpha$, and
of the coupling term between the tagged particle and the $\alpha$-th bath
oscillator, which contains the coupling function $F_{\alpha}(Q)$.

The new term

\begin{align}
H_E&=-ze\Phi(Q,t)-\sum_\alpha e_\alpha\Phi(X_\alpha,t)\notag\\
&=-E_0\sin{(\omega t)}(zeQ + \sum_\alpha e_\alpha X_{\alpha})
\end{align}

represents the influence of electric field on both the tagged particle (first
term on the RHS) and the bath oscillators (second term on the RHS). Here, $e$
is the unit charge and hence $ze$ is the total charge of tagged particle, where
e.g. $z=\pm 1$ for monovalent ions/electrons in a plasma or $z=-1$ for the case
of molecular dielectrics where a molecule polarizes into a negatively charged
electron cloud which oscillates about a positively charged molecular ion;
$e_\alpha$ is the net charge of bath particle $\alpha$ which is subjected to
the same polarization. Here, we only consider motion along the direction of
electric field.

In the Caldeira-Leggett Hamiltonian, the coupling function is taken to be
linear in the particle coordinate $Q$, $F_{\alpha}(Q)=c_{\alpha}Q$, where
$c_{\alpha}$ is known as the strength of coupling between the tagged atom and
the $\alpha$-th bath oscillator.

This Hamiltonian leads in a straightforward manner to the following system of
differential equations:
\begin{widetext}
\begin{align}
\frac{dQ}{dt}&=\frac{P}{m};\quad \frac{dP}{dt}=-V'(Q)+\sum_{\alpha}m_\alpha
c_\alpha(X_\alpha-\frac{c_\alpha Q}{\omega_\alpha^2})+zeE_0\sin{(\omega t)}\\
\frac{dX_\alpha}{dt}&=\frac{P_{\alpha}}{m_\alpha};
\frac{dP_\alpha}{dt}=-m_\alpha\omega_\alpha^2X_\alpha+m_\alpha c_\alpha
Q+e_\alpha E_0\sin{(\omega t)}\notag
\end{align}
\end{widetext}

From the second line, upon solving the second-order inhomogeneous ODE with the Green's function method (or 
alternatively with the Wronskian method), we get
\begin{align}
X_\alpha(t)&=X_\alpha(0)\cos{(\omega_\alpha
t)}+\frac{P_{\alpha}(0)\sin{(\omega_\alpha t)}}{m_\alpha\omega_\alpha}\notag\\
&+\int_{0}^t[c_{\alpha}Q(s)+\frac{e_\alpha}{m_\alpha} E_0\sin{(\omega
s)}]\frac{\sin{(\omega_\alpha(t-s))}}{\omega_\alpha}ds
\end{align}

The integral $\int_{0}^tc_\alpha
Q(s)\frac{\sin{(\omega_\alpha(t-s))}}{\omega_\alpha}ds$ can be evaluated via
integration by parts. Upon further denoting $E_\alpha(t)=\int_{0}^te_\alpha
E_0\sin{(\omega s)}\frac{\sin{(\omega_\alpha(t-s))}}{\omega_\alpha}ds$, we
obtain
\begin{align}
&X_\alpha(t)-\frac{c_\alpha
Q(t)}{\omega_\alpha^2}=\left(X_\alpha(0)-\frac{c_\alpha
Q(0)}{\omega_\alpha^2}\right)\cos{(\omega_\alpha t)}\notag\\
&+P_\alpha(0)\frac{\sin{(\omega_\alpha
t)}}{m_\alpha\omega_\alpha}-\int_{0}^t\frac{c_\alpha
P(s)\cos{(\omega_\alpha(t-s))}}{m\omega_\alpha^2}ds+\frac{E_\alpha(t)}{m_\alpha}
\end{align}

Substituting Eq. (5) into the equation for $P(t)$ in Eq. (3), we derive the
following Generalized Langevin Equation for the tagged particle motion, which accounts for the AC electric field acting on
both the tagged particle and the bath oscillators:
\begin{widetext}
\begin{align}
\frac{dP}{dt}&=-V'(Q(t))-\sum_\alpha\int_{0}^t\frac{m_\alpha\cos{(\omega_\alpha(t-s))}}{m\omega_\alpha^2}c_\alpha^2P(s)ds+zeE_0\sin{(\omega
t)}\notag\\
&+\sum_\alpha\{m_\alpha c_\alpha\left[X_\alpha(0)-\frac{c_\alpha
Q(0)}{\omega_\alpha^2}\right]\cos{(\omega_\alpha t)}+c_\alpha
P_\alpha(0)\frac{\sin{(\omega_\alpha t)}}{\omega_\alpha}+c_\alpha
E_\alpha(t)\}\notag\\
&=-V'(Q(t))-\int_{0}^t\nu(s)\frac{m_\alpha P(t-s)}{m}ds+zeE_0\sin{(\omega
t)}+F_P(t).
\end{align}
\end{widetext}

We have introduced the noise or stochastic force $F_P(t)$ which is equal to the
second line after the first equality,

\begin{align}
F_{p}(t)&=\sum_\alpha\{m_\alpha c_\alpha\left[X_\alpha(0)-\frac{c_\alpha
Q(0)}{\omega_\alpha^2}\right]\cos{(\omega_\alpha t)}\notag\\
&+c_\alpha P_\alpha(0)\frac{\sin{(\omega_\alpha t)}}{\omega_\alpha}+c_\alpha
E_\alpha(t)\}.
\end{align}

One should note that this expression for the stochastic force is identical with
the one derived by Zwanzig for a particle-bath system without external field,
except for the important term $c_\alpha E_\alpha(t)$, which is new and contains
the effect of the external AC field on the bath oscillators dynamics. This is a
crucial point because the dynamical response of the bath oscillators to the
external AC field in turn produces a modification of spectral properties of the
bath fluctuations, and thus leads to a new form of the stochastic force which
is different from those studied in previous works.

The memory function for the friction $\nu(t)=\sum_\alpha\frac{m_\alpha
c_\alpha^2}{m\omega_\alpha^2}\cos{(\omega_\alpha t)}$ is identical to the
memory function of systems without external time-dependent forces such as the
one derived by Zwanzig~\cite{Zwanzig}.

The integral in the function $E_\alpha(t)=\int_{0}^te_\alpha E_0\sin{(\omega
s)}\frac{\sin{(\omega_\alpha(t-s))}}{\omega_\alpha}ds$, can be evaluated using
trigonometric identities which leads to
\begin{equation}
E_\alpha(t)=\frac{e_\alpha E_0(\omega\sin{(\omega_\alpha
t)}-\omega_\alpha\sin{(\omega t)})}{\omega_\alpha(\omega^2-\omega_\alpha^2)}.
\end{equation}
As for the case without external time-dependent fields, we find that our
$F_P(t)$ is defined in terms of initial positions and momenta of bath
oscillators, but in our case it now also depends on the sinusoidal electric
field at time $t$. Note that, by shifting the time origin, it can be easily
verified that the statistical average is stationary. Following
Zwanzig~\cite{Zwanzig}, we assume the initial conditions for the bath
oscillators can be taken to be Boltzmann-distributed $\sim
\exp(-H_{B}/k_{B}T)$, where the bath is in thermal equilibrium with respect to
a frozen or constrained system coordinate $X(0)$.

Then for the average of $X$ and $P$ we find:
\begin{equation}
\langle X_\alpha(0)-\frac{c_\alpha Q(0)}{\omega_\alpha^2}\rangle=0, \quad
\langle P_\alpha(0)\rangle=0.
\end{equation}
The second moments are
\begin{equation}
\left\langle\left(X_\alpha(0)-\frac{c_\alpha
Q(0)}{\omega_\alpha^2}\right)^2\right\rangle=\frac{k_BT}{m_\alpha\omega_\alpha^2},
\quad \langle P_\alpha(0)^2\rangle=m_\alpha k_BT.
\end{equation}
Both these results are consistent with what one finds for systems without
external time-dependent fields, since obviously they descend directly from the
assumption of Boltzmann-distributed degrees of freedom at the initial time.

As for $\sum_\alpha c_\alpha E_\alpha(t)$, we first note that there is no
singularity when $\omega_\alpha\rightarrow\omega$ and
$\omega_\alpha\rightarrow0$:
\begin{align}
\lim_{\omega_\alpha\rightarrow\omega}\frac{\omega\sin{(\omega_\alpha
t)-\omega_\alpha\sin{(\omega
t)}}}{\omega_\alpha(\omega^2-\omega_\alpha^2)}&=-\frac{\omega t\cos{(\omega t
)}-\sin{(\omega t)}}{2\omega^2}\notag\\
\lim_{\omega_\alpha\rightarrow0}\frac{\omega\sin{(\omega_\alpha
t)-\omega_\alpha\sin{(\omega
t)}}}{\omega_\alpha(\omega^2-\omega_\alpha^2)}&=\frac{\omega t-\sin{(\omega
t)}}{\omega^2}.
\end{align}

Focusing on ions, atoms or molecules, $\omega_\alpha$ is at least in the THz
regime or higher, which is orders of magnitude larger than the frequency of the
applied field $\omega$ (this assumption of course may not hold for THz
spectroscopy, which deserves a separate treatment in future work). Hence, the
first term in the numerator in the RHS of Eq. (8) can be neglected along with the first term in the bracket in the denominator, and we can
approximate as follows:
\begin{equation}
\sum_\alpha c_\alpha E_\alpha(t)\approx \sum_{\alpha}\frac{c_\alpha
e_\alpha}{\omega_\alpha^2}E_0\sin{(\omega t)}\varpropto eE(t).
\end{equation}

We now take the Boltzmann average of the stochastic force, Eq. (7), and we
find:
\begin{equation}
\langle F_P(t)\rangle=\gamma eE(t),
\end{equation}
for some constant $\gamma$.
That is, the average of the stochastic force $F_P$ is not zero, unlike
in all previously studied GLEs of particle-bath systems and, remarkably, is
directly proportional to the AC field.

Now, by direct calculation, using Eqs. (9) and (10) and standard
trigonometric identities, we can get the fluctuation-disspation theorem (FDT)
for our particle-bath Hamiltonian under a uniform AC field:
\begin{widetext}
\begin{align}
\langle F_P(t)F_P(t')\rangle&=\frac{1}{Z_N}\int
F_P(t)F_P(t')\exp{\left(-\frac{H_B}{k_BT}\right)}d\mathbf{X}(0)
d\mathbf{P}(0)\notag\\
&=\sum_{\alpha}\left(m_\alpha
c_\alpha^2\frac{k_BT}{\omega_\alpha^2}\cos{(\omega_\alpha
t)}\cos{(\omega_\alpha t')}+m_\alpha
c_\alpha^2\frac{k_BT}{\omega_\alpha^2}\sin{(\omega_\alpha
t)}\sin{(\omega_\alpha t')}\right)+(\gamma e)^2E(t)E(t')\notag\\
&=k_BT\sum_{\alpha}\frac{m_\alpha
c_\alpha^2}{\omega_\alpha^2}\cos{(\omega_\alpha(t-t'))}+(\gamma
e)^2E(t)E(t')\notag\\
&=mk_BT\nu(t-t')+(\gamma e)^2E(t)E(t')
\end{align}
\end{widetext}
where $Z_N$ is the canonical partition function
\begin{equation}
Z_N=\int\exp{(-\frac{H_B}{k_BT})}d\mathbf{X}(0)d\mathbf{P}(0)
\end{equation}
and $\mathbf{X}(0)=\{X_1(0),X_2(0),...\},~\mathbf{P}(0)=\{P_1(0),P_2(0),...\}$.

Equation (14) is a key result of this paper and is the FDT associated with the
GLE given by Eq. (6).
This is a remarkable result which shows that in the presence of an external AC
field which affects the microscopic dynamics of both the tagged particle and
the bath oscillators, the strength of the noise is no longer proportional to
thermal energy only, but also has an important deterministic contribution
proportional to the AC field amplitude squared.

The implications of this result for statistical mechanics and its applications
in condensed matter are vast and profound and will be explored in future works,
including experimental verifications and extension to nonequilibrium enviornments~\cite{Harada,Sasa,Maes,Steffenoni}.
We can however notice that the presence of the external AC field makes it
impossible for the noise to be white noise. This is true even if the memory
function is delta-correlated, because of the second term controlled by the AC
field in Eq. (14), which inevitably introduces a deterministic
temperature-independent correlation into the noise.

In conclusion, we have introduced a more general version of the classical
particle-bath Hamiltonian, which is used as a starting point to derive
Generalized Langevin Equations, for systems subjected to an external
time-dependent (oscillating) field. Unlike in previous models where the bath
oscillators were always taken to be unaffected by the external field, here we added the
time-dependent force due to the field to both the Hamiltonian of the particle
and the Hamiltonian of the bath oscillators. The resulting Hamiltonian has been
solved analytically and the resulting GLE and fluctuation-dissipation theorem
have been found. The formal structure of the GLE is still formally identical to
that of standard GLE with external field acting on the particle
only~\cite{Kubo} (and the memory function for the friction is the same), but
the stochastic force is very different. Its ensemble average is
non-zero and directly proportional to the AC field. The associated
fluctuation-dissipation theorem, remarkably, has an additional term given by the
time-correlation of the AC field, and is thus quadratic in the field amplitude.

An immediate application of this result is to dielectric relaxation and
dielectric spectroscopy of liquids and glasses. The Debye model treats each
molecule as fully independent from all other molecules in the material and
describes it with a Langevin equation for the orientation of the molecule in
the field~\cite{Debye,Coffey}. More refined models, e.g. Mode-Coupling Theory,
are able to account for the coupling of each molecule to collective degrees of
freedom~\cite{Goetze,Blochowitz}, but do not resolve the combined effect of the
AC field on the dynamics of both the single molecule and the other molecules
which provide the memory effect in dielectric relaxation. The GLE derived here
will open the possibility of describing both these effects at the same time,
within the GLE picture that has been proposed more recently for the dielectric response of liquids and glasses~\cite{Rubi1,Rubi2,Cui}.
Furthermore, the GLE derived here can be used as the starting point for a more
microscopic description of nonlinear effects in dielectric relaxation of
supercooled liquids under strong fields~\cite{Loidl,Richert}.

In the context of microrheology of soft and biological matter, the FDT is the starting point to derive the key relationship (generalized Stokes-Einstein relation, GSE) that is used to infer the viscoelastic response of the medium to an oscillatory mechanical field from the optically monitored dynamics of a probe particle~\cite{Weitz,MacKintosh,Lubensky,Bausch,Cicuta}. This method based on FDT has found widespread application in biophysics, to measure the viscoelasticity of cytoskeleton and other biological matters. 
However, all these works, including the most recent ones (see e.g. Ref.~\cite{Kilfoil}) always neglect the effect of the oscillatory field on the dynamics of the medium (the bath).
Clearly, the system as a whole is composed of many building blocks (molecules, filaments) which also respond dynamically to the oscillatory shear field and behave as the “bath” for the probe particle. Therefore, the same effect described in our paper is certainly present also in all those systems, and yet it has always been neglected in all the literature so far. 

Further applications of the proposed framework include quantum dissipative
transport and Josephson tunnelling with dissipation~\cite{Breuer,Caldeira,Zwerger},
driven dynamics of colloids in soft matter systems~\cite{Voigtmann,
Barrat,Yurchenko}, molecular dynamics simulations of
liquids~\cite{Schilling} and of amorphous solids in oscillatory
shear~\cite{Damart}.

B. Cui acknowledges the financial support from the CSC-Cambridge scholarship.

\end{document}